\documentclass[pra,aps,amsfonts,amsmath,superscriptaddress,twocolumn,showpacs,floatfix]{revtex4-1}
\usepackage{graphicx} 
\usepackage{epsfig} 

\begin{document}

\title{How atomic nuclei cluster}

\author{J.-P. Ebran}
\affiliation{CEA/DAM/DIF, F-91297 Arpajon, France}
\author{E. Khan}
\affiliation{Institut de Physique Nucl\'eaire, Universit\'e Paris-Sud, IN2P3-CNRS, F-91406 Orsay Cedex, France}
\author{T. Niksic}
\author{D. Vretenar}
\affiliation{Physics Department, Faculty of Science, University of
Zagreb, 10000 Zagreb, Croatia}

\begin{abstract}
Nucleonic matter displays a quantum liquid structure, but in some
cases finite nuclei behave like molecules composed of clusters of
protons and neutrons. Clustering is a recurrent feature in light
nuclei, from beryllium to nickel. For instance, in $^{12}$C the Hoyle
state, crucial for stellar nucleosynthesis, can be described as a
nuclear molecule consisting of three alpha-particles. The mechanism of
cluster formation, however, has not yet been fully understood. We show
that the origin of clustering can be traced back to the depth of the
confining nuclear potential. By employing the theoretical framework of
energy density functionals that encompasses both cluster and quantum
liquid-drop aspects of nuclei, it is shown that the depth of the
potential determines the energy spacings between single-nucleon
orbitals, the localization of the corresponding wave functions and,
therefore, the degree of nucleonic density clustering. Relativistic
functionals, in particular, are characterized by deep single-nucleon
potentials. When compared to non-relativistic functionals that yield
similar ground-state properties (binding energy, deformation, radii),
they predict the occurrence of much more pronounced cluster
structures. More generally, clustering is considered as a transitional
phenomenon between crystalline and quantum liquid phases of fermionic
systems. 
\end{abstract}


\date{\today}

\maketitle

The occurrence of molecular states in atomic nuclei and the formation
of clusters of nucleons were already predicted in the 30's by von
Weizs\"acker and Wheeler \cite{1,2}. Even though the description of nuclear
dynamics became predominantly based on the concept of independent
nucleons in a mean-field potential, a renewed interest in clustering
phenomena in the 60's led to the development of dedicated
theoretical methods \cite{3}. Numerous experimental studies have revealed a
wealth of data on clustering phenomena in light nuclei \cite{4}, and modern
theoretical approaches use microscopic models that fully take into
account single-nucleon degrees of freedom \cite{5,6,7}. Clustering gives rise to
nuclear molecules. For instance, in $^{12}$C the second 0$^+$ state the
Hoyle state that plays a critical role in stellar nucleosynthesis, is
predicted to display a three-$\alpha$ structure \cite{8,9}. The binding energy of the
$\alpha$-particle, formed from two protons and two neutrons, is much larger
than in other light nuclei. Cluster radioactivity \cite{10}, discovered in the
80's, is another manifestation of clustering in atomic nuclei.
Experimental signatures of clustering are usually indirect.
Quasi-molecular resonances are probed by scattering one cluster on
another, such as in the $^{12}$C+$^{12}$C system \cite{4,11}, and cluster structures are
also discernible in the breakup of nuclei. Evidence has been reported
for the formation of clusters in ground and excited states of a number
of $\alpha$-conjugate nuclei \cite{4}, that is nuclei with an equal even number of
protons and neutrons, from $^8$Be to $^{56}$Ni.

The mechanism of cluster formation in nuclei has not yet been fully
understood. Deformation plays an important role because it removes the
degeneracy of single-nucleon levels associated with spherical
symmetry. At specific deformations the shell structure can restore
degeneracies corresponding, for instance, to a 2:1 ratio of the large
axis over the small axis of a quadrupole deformed system4.
Consequently, the restored degeneracy of deformed shell closures
facilitates the formation of clusters. However, this is a rather
qualitative explanation because clustering phenomena cannot generally
be explained by accidental degeneracies. Clustering is an essential
feature of many-nucleon dynamics that coexists with the nuclear
mean-field. Therefore, although in most cluster models the existence
of such structures is assumed a priori and the corresponding effective
interactions are adjusted to the binding energies and scattering phase
shifts of these configurations, a fully microscopic understanding of
cluster formation necessitates a more general description that
encompasses both cluster and quantum liquid-drop aspects in light as
well as in heavier nuclei. The aim of this work is to address the
origin of clustering, i.e. to find out the conditions for cluster
formation in ground states of finite nuclei, starting from a fully
microscopic description based on the framework of energy density
functionals.

At present the only comprehensive approach to nuclear structure is
based on the framework of energy density functionals (EDFs). Nuclear
EDFs enable a complete and accurate description of ground-state
properties and collective excitations over the whole nuclide chart
\cite{12,13,14}. In practical implementations nuclear EDFs are analogous to
Kohn-Sham Density Functional Theory, the most widely used method for
electronic structure calculations in condensed matter physics and
quantum chemistry. In the nuclear case the many-body dynamics is
represented by independent nucleons moving in a local self-consistent
mean-field potential that corresponds to the actual density and
current distribution of a given nucleus. Both non-relativistic and
relativistic realizations of EDFs are employed in studies of nuclear
matter and finite nuclei. A nuclear EDF is universal in the sense
that, for a given inter-nucleon interaction, it has the same
functional form for all systems. Using a small set of global
parameters adjusted to empirical properties of homogeneous nuclear
matter and data on finite nuclei, a universal functional provides a
description of the structure of nuclei across the chart of nuclides
and, therefore, describes the coexistence of cluster and
quantum-liquid aspects of light nuclei. 

A number of recent studies based on nuclear EDFs or the mean-field
approach have analysed cluster structures in $\alpha$-conjugate nuclei
\cite{15,16,17,18,19,20,21}. In Fig. 1 we display the self-consistent ground-state 
densities of $^{20}$Ne, calculated with two widely used functionals
that are representative for the two classes of nuclear EDFs: the
non-relativistic Skyrme SLy4 \cite{22}, and the relativistic functional
DD-ME2 \cite{23}. The equilibrium shape of $^{20}$Ne is a prolate, axially
symmetric quadrupole ellipsoid. Although they have not been
specifically adjusted to this mass region, both functionals reproduce
the empirical ground-state properties of this nucleus: the
experimental binding energy 160.6 MeV, the radius of the proton
distribution 2.90 fm \cite{24}, and the radius of the matter distribution
2.85 fm \cite{25}, with a typical accuracy of ~1\%. It is remarkable that,
although these functionals predict similar values for the binding
energy, charge and matter radii, as well as the quadrupole 
deformation of equilibrium shape of $^{20}$Ne, yet the corresponding
single-nucleon densities are qualitatively very different. The density
calculated with SLy4 displays a smooth behaviour characteristic for a
Fermi liquid, with an extended surface region in which the density
very gradually decreases from the central value of ~ 0.16 fm$^{-3}$. The
relativistic functional DD-ME2, on the other hand, predicts an
equilibrium density that is much more localized. The formation of
cluster structures is clearly visible, with density spikes as large as
~ 0.2 fm$^{-3}$, and a much narrower surface region. 

\begin{figure}[tb]
\begin{center}
\scalebox{0.05}{\includegraphics{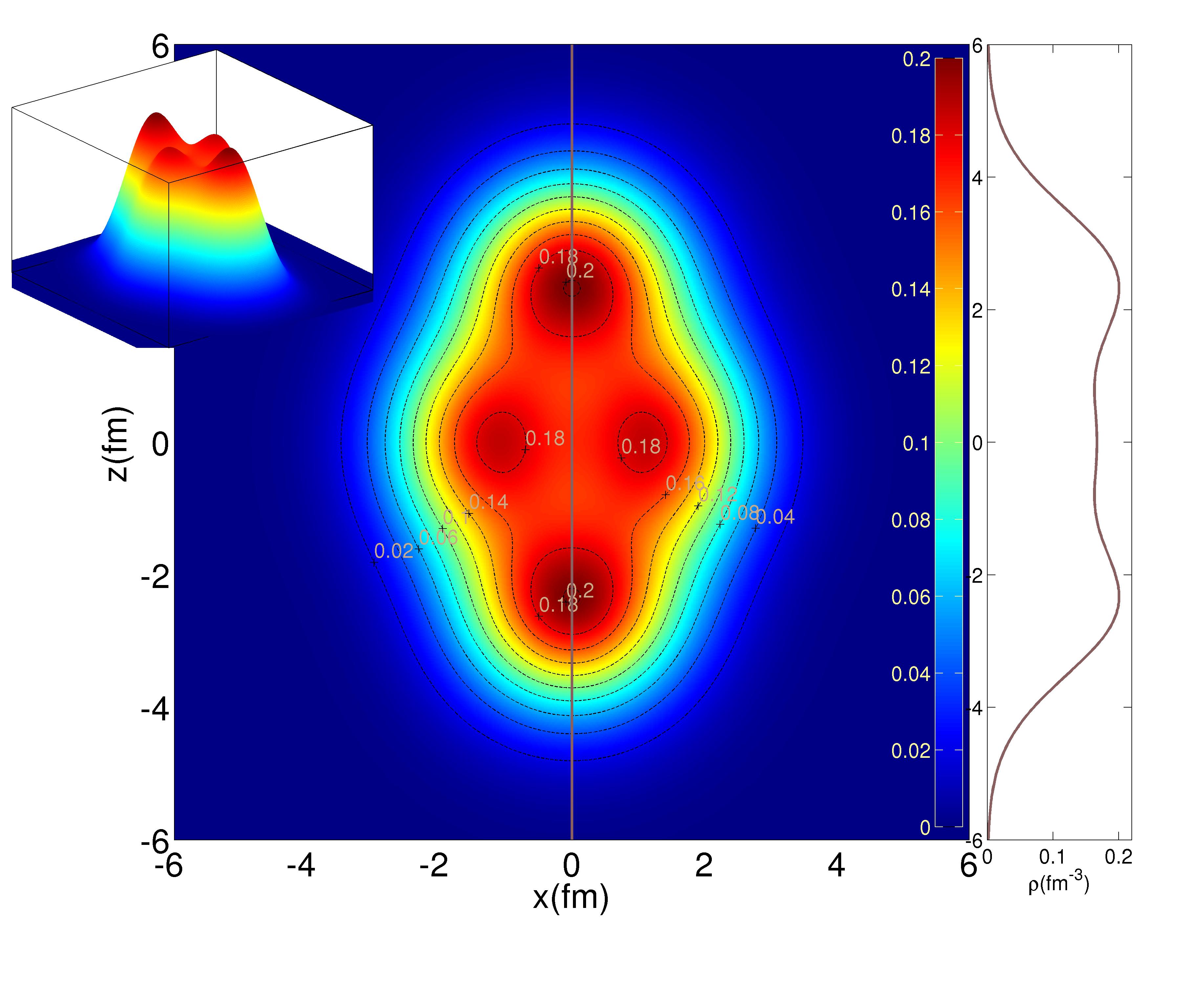}}
\scalebox{0.05}{\includegraphics{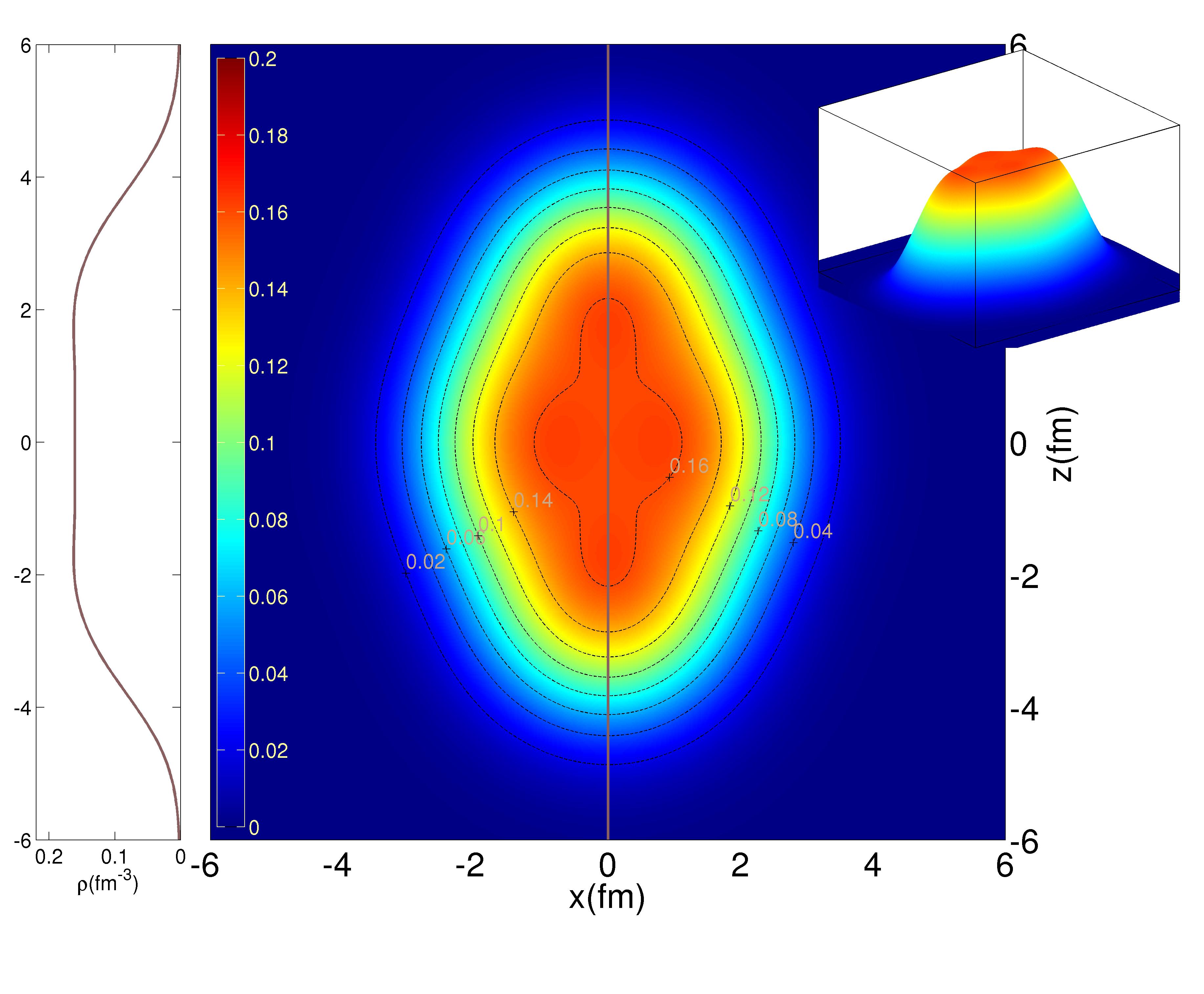}}
\caption{Plots of self-consistent ground-state
densities of $^{20}$Ne, calculated with the nuclear energy density
functionals: DD-ME2\cite{23} (top), and Skyrme SLy4 \cite{22,30} (bottom). The
densities (in units of fm$^{-3}$) are plotted in the x-z plane of the
intrinsic frame of reference that coincides with the principal axes of
the nucleus, with z chosen as the symmetry axis. The inserts show the
corresponding three-dimensional density plots and the density profiles
along the symmetry axis, respectively.}
\label{fig:dens}
\end{center}
\end{figure}

Understanding the difference in the equilibrium densities of $^{20}$Ne
calculated with SLy4 and DD-ME2 is the key to the mechanism of cluster
formation in this mass region of $\alpha$-conjugate deformed nuclei. The
axially symmetric deformation of the nuclear mean-field removes the
degeneracy of spherical single-nucleon levels, and nucleons pairwise
(spin up and down) occupy orbitals characterized by time-reversal
degeneracy. For large deformations these levels can be labelled by a
set of asymptotic Nilsson quantum numbers\cite{26} and, because of the
relatively weak Coulomb interaction in light nuclei, the localization
of proton and neutron orbitals is similar in Z = N nuclei. In the
specific case of $^{20}$Ne, ten protons and ten neutrons occupy five
deformed Nilsson levels, with the energy spacing between these levels
being proportional to the deformation of the single-nucleon potential.
Figure 2 shows the partial single-nucleon densities that correspond to
highest occupied Nilsson orbital. Even without introducing a
quantitative measure of localization, it is obvious that DD-ME2
predicts a much more localized density distribution. Similar results,
i.e. more localized density distributions calculated with DD-ME2, are
also obtained for the other four occupied orbitals.

\begin{figure}[tb]
\begin{center}
\scalebox{0.05}{\includegraphics{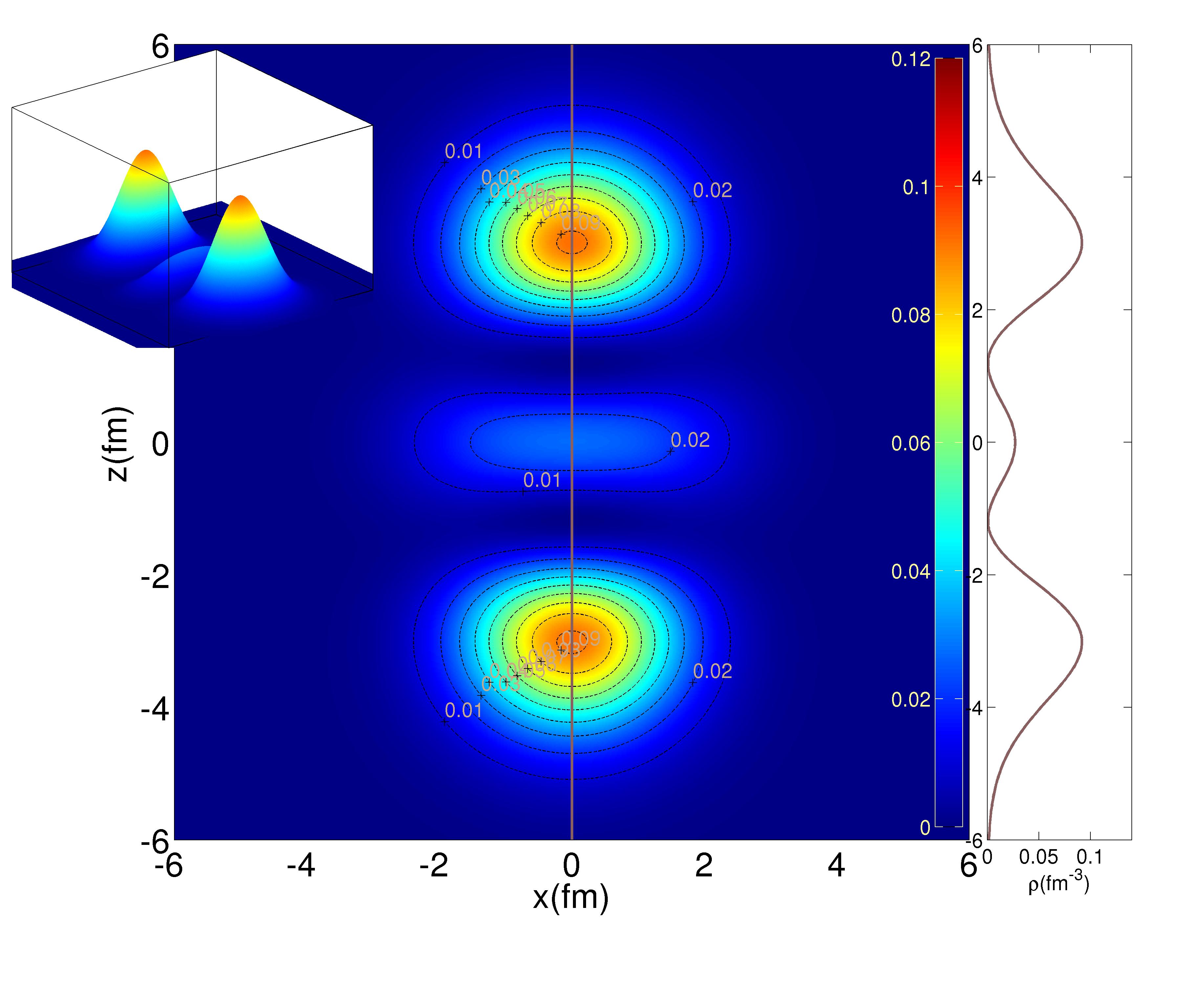}}
\scalebox{0.05}{\includegraphics{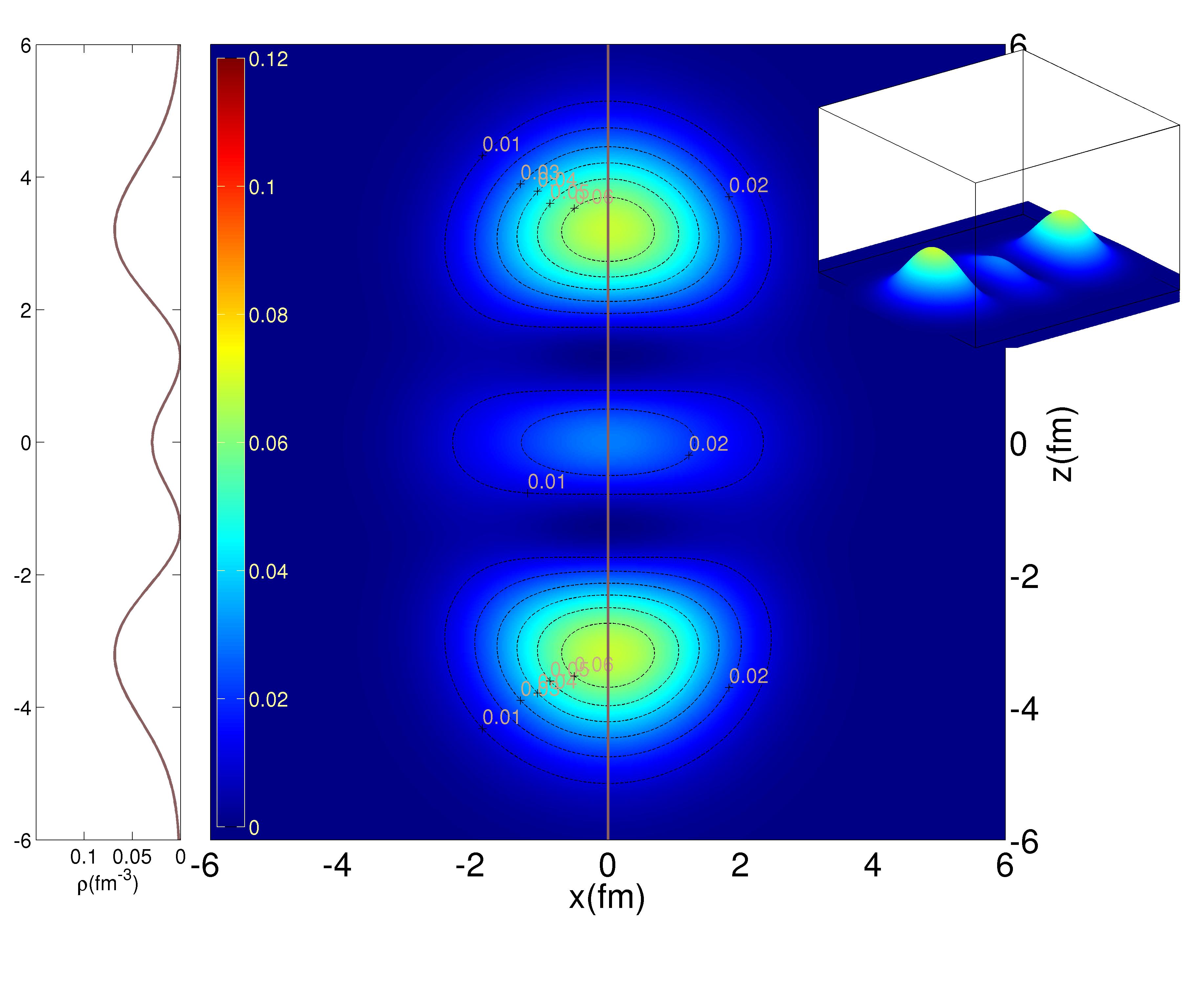}}
\caption{Partial nucleon density distributions that correspond to 
highest occupied (2 protons spin up and down, and 2 neutrons spin up
and down) in $^{20}$Ne: $\Omega^\pi$[Nnz$\Lambda$] = 1/2$^+$[220], calculated with the nuclear
energy density functionals: DD-ME2\cite{23} (top), and
SLy4\cite{22,30} (bottom).}
\label{fig:dens}
\end{center}
\end{figure}

Localization of densities that correspond to single-particle orbitals
is a necessary precondition for the formation of clusters, and this
effect can be traced back to the corresponding single-nucleon spectra.
The comparison of spectra calculated with the two functionals shows
that the one obtained with DD-ME2 is more spread out, and the more
pronounced energy spacings between single-particle levels are also
reflected in the more localized wave functions and partial densities.
Starting from degenerate spherical single-particle levels, the
splitting of the corresponding Nilsson deformed states is proportional
to the deformation and to the depth of the potential. Since the two
functionals predict almost identical equilibrium deformations and
radii for $^{20}$Ne, the different energy spacings in the single-nucleon
spectra must reflect the difference in the corresponding potentials.
In fact, the self-consistent mean-field potential of DD-ME2 is
considerably deeper than that of SLy4. In the centre of the nucleus
the depth of the DD-ME2 single-neutron potential is -78.6 MeV, whereas
the depth of the SLy4 potential is -69.5 MeV. The corresponding values
of the single-proton potentials are: -72.8 MeV for DD-ME2, and -64.6
MeV for SLy4. The effect of the potential depth on the localization of
wave functions is schematically illustrated in Fig. 3 where, as an
approximation to nuclear potentials, we plot three harmonic oscillator
potentials with different values of the depth: 30, 45 and 60 MeV, but
with the same radius R = 3 fm. The radial wave functions of the
corresponding p-states are shown in the lower panel. The oscillator
length b determines the position of the maximum and the dispersion of
the wave function\cite{27}. The deeper the potential, the smaller the
oscillator length, and the wave functions become more localised. At
the origin of clustering is, therefore, the depth of the
self-consistent single-nucleon mean-field potential associated with a
given energy density functional. By performing a series of
self-consistent mean-field calculations using a variety of
non-relativistic and relativistic functionals, not only for $^{20}$Ne, but
also for $^{24}$Mg, $^{28}$Si and $^{32}$S, we have verified that pronounced cluster
structures in deformed equilibrium shapes indeed occur only for deep
single-nucleon potentials.

\begin{figure}[tb]
\begin{center}
\scalebox{0.05}{\includegraphics{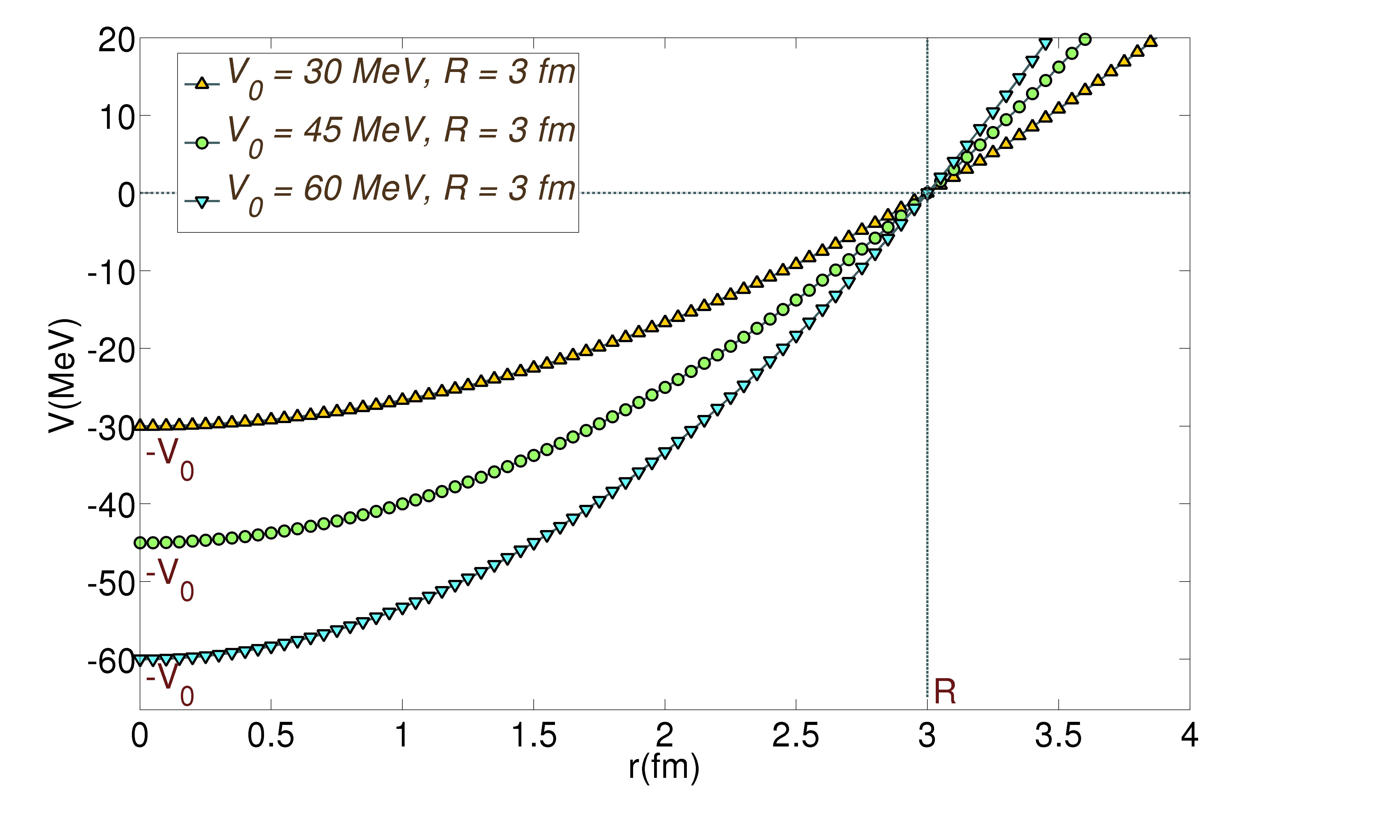}}
\scalebox{0.05}{\includegraphics{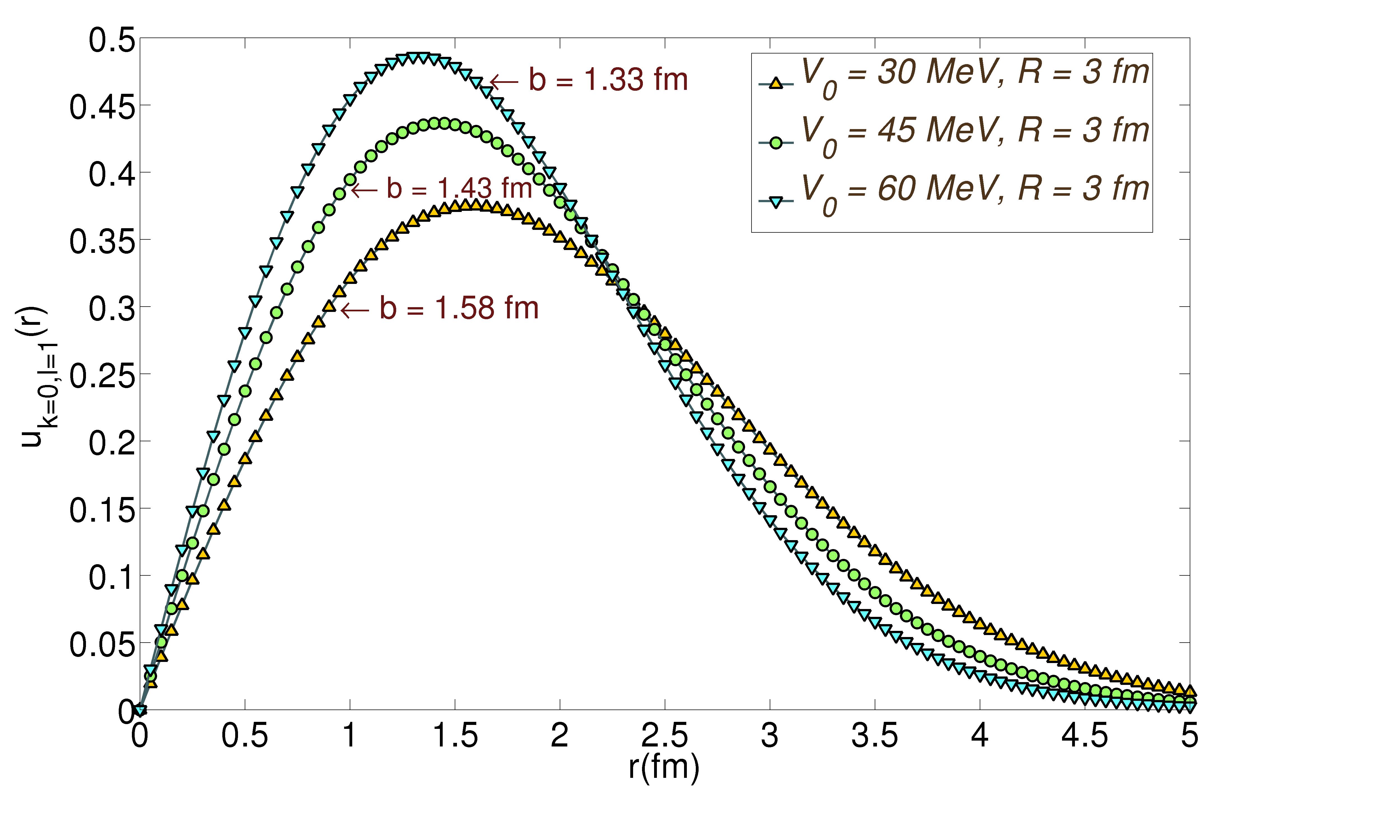}}
\caption{Top: Harmonic oscillator potentials for three different 
values of the depth: 30, 45 and 60 MeV, with the same radius R = 3 fm.
Bottom: the radial wave functions of the corresponding first p-state.
The position of the maximum is determined by the oscillator length b.}
\label{fig:dens}
\end{center}
\end{figure}

The difference of the potential depths calculated with DD-ME2 and SLy4
is characteristic for relativistic vs non-relativistic self-consistent
potentials. The depth of a relativistic potential is determined by the
difference of two large fields: an attractive (negative) Lorentz
scalar potential of magnitude ~400 MeV, and a repulsive Lorentz vector
potential ~320 MeV (plus the repulsive Coulomb potential for
protons) \cite{12,13,14}. The choice of these potentials is further constrained
by the fact that their sum ($\sim$ 700 MeV) determines the effective
single-nucleon spin-orbit potential. In a non-relativistic approach
the spin-orbit potential is included in a purely phenomenological way,
with the strength of the interaction adjusted to empirical energy
spacings between spin-orbit partner states. Since the relativistic
scalar and vector fields determine both the effective spin-orbit
potential and the self-consistent single-nucleon mean-field, for all
relativistic functionals the latter is found to be deeper than the
non-relativistic mean-field potentials, for which no such constraint
arise. 

More generally, fermionic systems can exhibit a crystalline phase or,
on the other extreme, a quantum liquid phase. B. Mottelson considered
the quantality \cite{28} parameter to show that nuclear matter
displays a quantum liquid structure. This concept can be generalised
by considering nuclear clusters as transitional states between
crystalline and quantum liquid phases (Fig. 4). The dimensionless
ratio $\alpha$ = b/r$_0$ , where b is the dispersion of the nucleon
wave-function, and r$_0$ the typical inter-nucleon distance ($\sim$ 1.2 fm), is
the natural parameter to quantify nuclear clustering, in analogy with
similar considerations in condensed matter\cite{29}. 

For a harmonic oscillator 

\begin{equation}
\alpha=\frac{\sqrt{\hbar R}}{r_0(2mV_0)^{1/4}} 
\end{equation}

where V$_0$ is the
depth of the potential, and R the radius of the system.

When $\alpha$ $>$1 nucleons are
delocalised and the nucleus has a quantum liquid structure. The
transition to a cluster state occurs when $\alpha$ $\sim$ 1, nucleons become more
localised and form a molecular structure (Fig. 4). In the present
analysis we find that $\alpha$ $<$ 1 for the relativistic functional, whereas
$\alpha$ $>$1 for the non-relativistic functional. Moreover, from its
definition in the case of a harmonic oscillator potential (see Eq. (1)), $\alpha$ obviously increases with the number of nucleons (nuclear
radius). Cluster states, therefore, are less likely to appear in
heavier nuclei.

\begin{figure}[tb]
\begin{center}
\scalebox{0.4}{\includegraphics[trim=3cm 7cm 0cm 7cm]{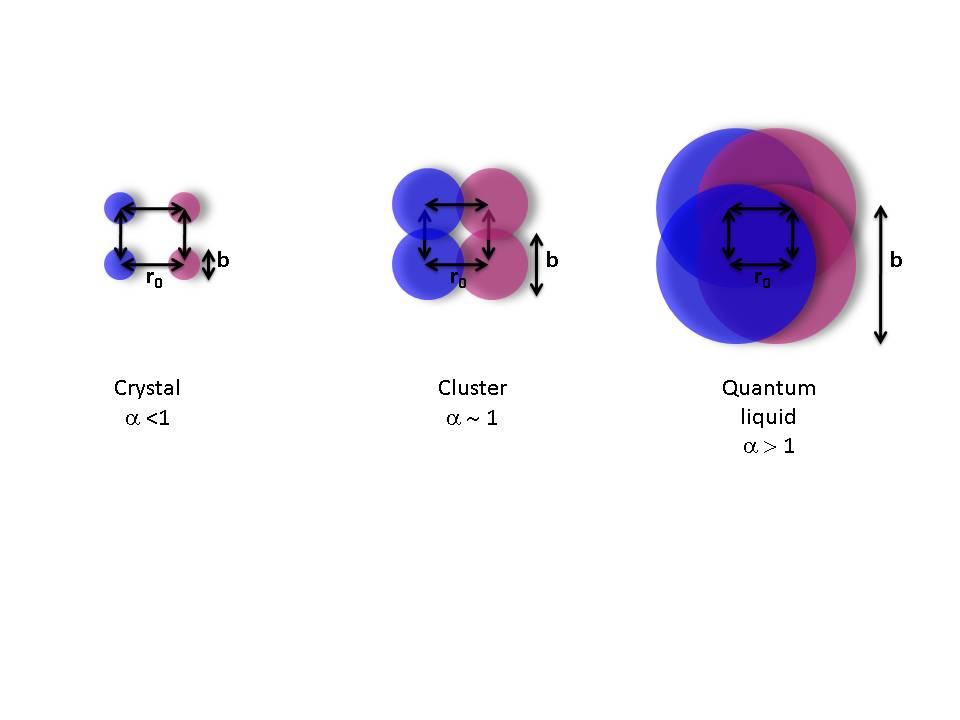}}
\caption{Transition from crystalline to a quantum liquid phase, 
including the cluster phase. $\alpha$ = b/r$_0$ , where b is the dispersion of
the fermion wave-function, and r$_0$ the typical inter-fermion distance.}
\label{fig:dens}
\end{center}
\end{figure}

\end{document}